\documentclass[aps,prb,twocolumn,superscriptaddress]{revtex4-1}

\usepackage{amsmath, amssymb}
\usepackage{graphicx, hyperref}
\usepackage{color}

\begin{document}

\title{From atomic layer to the bulk: low-temperature atomistic structure,\\ferroelectric and electronic properties of SnTe films}

\author{Thaneshwor P. Kaloni}
\affiliation{Department of Physics, University of Arkansas, Fayetteville, AR 72701, USA}

\author{Kai Chang}
\affiliation{Max-Planck Institute of Microstructure Physics, Halle 06120, Germany}
\affiliation{State Key Laboratory of Low-Dimensional Quantum Physics, Department of Physics, Tsinghua University, Beijing 100084, China}
\affiliation{Collaborative Innovation Center of Quantum Matter, Beijing 100084, China}

\author{ Brandon J. Miller}
\affiliation{Department of Physics, University of Arkansas, Fayetteville, AR 72701, USA}

\author{Qi-Kun Xue}
\affiliation{State Key Laboratory of Low-Dimensional Quantum Physics, Department of Physics, Tsinghua University, Beijing 100084, China}
\affiliation{Collaborative Innovation Center of Quantum Matter, Beijing 100084, China}

\author{Xi Chen}
\affiliation{State Key Laboratory of Low-Dimensional Quantum Physics, Department of Physics, Tsinghua University, Beijing 100084, China}
\affiliation{Collaborative Innovation Center of Quantum Matter, Beijing 100084, China}

\author{Shuai-Hua Ji}
\affiliation{State Key Laboratory of Low-Dimensional Quantum Physics, Department of Physics, Tsinghua University, Beijing 100084, China}
\affiliation{Collaborative Innovation Center of Quantum Matter, Beijing 100084, China}
\affiliation{RIKEN Center for Emergent Matter Science (CEMS), Wako, Saitama 351-0198, Japan}

\author{Stuart S. P. Parkin}
\affiliation{Max-Planck Institute of Microstructure Physics, Halle 06120, Germany}

\author{Salvador Barraza-Lopez}
\email{sbarraza@uark.edu}
\affiliation{Department of Physics, University of Arkansas, Fayetteville, AR 72701, USA}
\affiliation{Institute of Nanoscale Science and Engineering, University of Arkansas, Fayetteville, AR 72701, USA}

\date{\today}

\begin{abstract}
SnTe hosts ferroelectricity that competes with its weak non-trivial band topology: in the high-symmetry rocksalt structure --in which its intrinsic electric dipole is quenched-- this material develops metallic surface bands, but in its rhombic ground-state configuration --that hosts a non-zero spontaneous electric dipole-- the crystalline symmetry is lowered and the presence of surface electronic bands is not guaranteed. Here, the type of ferroelectric coupling and the atomistic and electronic structure of SnTe films ranging from 2 to 40 ALs are examined on freestanding samples, to which {atomic layers were gradually added}. 4 AL SnTe films are antiferroelectrically-coupled, while thicker freestanding SnTe films are ferroelectrically-coupled. The electronic band gap reduces its magnitude in going from 2 ALs to 40 ALs but it does not close due to the rhombic nature of the structure. These results bridge the structure of SnTe films from the monolayer to the bulk.
\end{abstract}

\maketitle

\section{Introduction}\label{sec:1}

IV-VI compounds can form bulk rocksalt, orthorhombic or rhombic  ground state structures depending on their average atomic number:\cite{Littlewood1} PbS is a textbook example of rocksalt structure\cite{Kittel1} that lacks an electric dipole, SnSe is an orthorhombic layered compound with antiferroelectric coupling (labeled $AB$) among successive layers,\cite{Gangjia,Dravid2} and SnTe develops a ferroelectric coupling (labeled $AA$) on its rhombic phase. {
Bulk SnTe is a well studied material that nevertheless continues to provide new Physical phenomena. Studies exist of its optical and electronic properties \cite{Vago_1946,Brebrick_1962,Damon_1963,Zamel_1965,Riedl_1966} that include magnetoresistance\cite{Burke_1965,Savage_1972,Allgaier_1972,Okazaki_2018}, the influence of temperature on electron transport\cite{Burke_1969}, the evolution of  theoretical \cite{Lin_1967,Rogers_1968,Tung_1969,Rabe_1985} and experimental \cite{Tsu_1968,Kemeny_1976,Littlewood_2010} electronic band structure methodologies, the relation of carrier concentration and anomalous resistivity with the rhombic to rocksalt phase transition \cite{Kobayashi_1976,Grassie_1979,Katayama_1980}, and superconductivity \cite{Allen_1969,Hein_1969}.

Additionally, experimental studies of structural phase transitions on these {\em diatomic ferroelectrics} were performed with M\"osbauer \cite{Keune_1974} and Raman spectroscopies \cite{Brillson_1974,Shimada_1977}, neutron scattering\cite{Pawley_1966,Scott_1974,Iizumi_1975,Bevolo_1976,Knox_2014,Li_2014} and x-ray photoemission \cite{Shalvoy_1977,Knox_2014,Neill_2017}. Theories that explain such transitions based on lattice dynamics have been developed\cite{Cowley_1969,Natori_1976,Littlewood_1980,Bilz_1982,Strauch_1987,Cowley_1996} with an emphasis on soft-phonon modes \cite{Gillis_1969,Gillis_1972} and the corresponding softening of elastic constants \cite{Beattie_1969}. The combination of a temperature-dependent thermal \cite{Damon_1966} and electronic conductivities \cite{Burke_1969,Kobayashi_1976,Grassie_1979,Katayama_1980} make SnTe a model thermoelectric material\cite{Li_2014,Li_2014b}. The rhombic structure and ferroelectric ordering of SnTe occurs at temperatures below 150 K with a lattice constant $a_0=6.325$ \AA{} and a rhombic angle $\alpha=89.895^{\circ}$ \cite{Littlewood1,book1,doi:10.1143/JPSJ.38.443,Littlewood2,Sugai,Jantsch1983,Brillson} among lattice vectors.

A twist on recreating the parity anomaly by electronic band inversion on the group-IV-VI material family\cite{Fradkin_1986,Pankratov_1987} culminated on a rediscovery of SnTe as a topological crystalline insulator later on \cite{Hsieh_2012,Tanaka_2012}. But having a rhombic symmetry, { i.e., a lower symmetry} than that of a rocksalt structure, its surface electronic states along the (100) direction must be gapped at low temperature\cite{PhysRevB.90.161108}. Nowadays, the \emph{coupling} among the temperature-dependent degrees of freedom discussed in previous paragraphs-- electronic band structure\cite{Li_2014c}, thermoelectricity\cite{Rameshti_2016}-- as well as discoveries of higher-order topology\cite{Schindler_2018} continue to find their way to thick SnTe slabs.

}


{ At the same time, interest on ultrathin SnTe originated due to theoretical predictions of ferroelectricity on these films \cite{singh_apl_2014_ges_gese_sns_snse,Mehboudi2016,huang_jcp_2016_ges_gese_sns_snse,kamal_prb_2016_iv_vi_monolayers,wu_prb_2017_all,liu_prl_2018_snte,slawinska_2d_mat_2018_mmls}
and their experimental fabrication \cite{Chang274,advmat}. These} SnTe slabs have not been created by capping a bulk sample, but grown from the bottom up\cite{Chang274,Stroscio,advmat}. { And} while common theoretical approaches assume a slab can be obtained by cutting two opposing surfaces of bulk rocksalt\cite{Hsieh2012} or rhombic\cite{PhysRevB.90.161108,arxiv1,liu_prl_2018_snte} bulk samples, the present work is aimed to explore the structural evolution of a freestanding SnTe slab containing $2n$ ALs by the successive addition of a 2 ALs \emph{ in the overall lowest-energy conformation} to the slab containing $2(n-1)$ ALs with $n$ a positive integer, complementing the experimental results of Ref.~\onlinecite{advmat}.

{ Toward this goal, the following points will be established here}: (a) 4 AL SnTe films are antiferroelectrically coupled, while thicker suspended SnTe films turned out to be ferroelectrically coupled. (b) In going from 2 to 40 ALs, the rhombic angle $\Delta \alpha$, defined as $90^{\circ}-\alpha$, decreases from about $2^{\circ}$ down to $\sim 0.11^{\circ}$, which is close to its experimental magnitude in the bulk. The manuscript is structured as follows. Technical details are provided in Sec.~\ref{sec:methods}, followed by results and discussion in Sec.~\ref{sec:results} and conclusions  in Sec.~\ref{sec:conclusions}.

\section{Technical details}\label{sec:methods}

We performed \textit{ab-intio} calculations with the {\em SIESTA} code\cite{siesta} (that employs localized numeric atomic orbitals\cite{Junquera} and norm-conserving Troullier-Martins pseudopotentials\cite{tm}) with van der Waals corrections of the Berland-Per Hyldgaard (BH) type\cite{PhysRevB.90.075148} { (and also known as \emph{cx-vdW-DF1})} as implemented by Rom\'an-P\'erez and Soler,\cite{soler} on pseudopotentials whose radii were optimized in-house.\cite{rivero} The real-space grid in which the Poisson equation is solved has a cutoff energy of 300 Ry. A Monkhorst-Pack\cite{mp} mesh of $18\times18\times1$ $k-$points was employed { in calculations involving unit cells, and a $3\times3\times1$ $k-$point mesh for calculations on 11$\times$11 supercells containing vacancies}. Standard (DZP) basis sets\cite{Junquera} with a PAO.Energyshift flag of 0.002205 Ry were used. The vertical vacuum among periodic slabs was set to 60 \AA{}, and dipole corrections were turned on. Structural optimizations were performed with a force tolerance of $10^{-3}$ eV/\AA. Spin-orbit coupling was turned on only after the electronic structure was optimized. Simulated STM images\cite{PhysRevLett.50.1998,PhysRevB.37.8327} were obtained by adding the electronic densities of individual wave functions { in an energy window consistent with experimental energy ranges}, and captured { 2.5} \AA{} above the SnTe film (see Refs.~\onlinecite{nlgr1,aplgr2,PhysRevB.91.115413}).

In addition, ultra-thin SnTe films were grown on 6H-SiC(0001) substrates that were sublimated to host epitaxial graphene layers\cite{Berger} { employing} substrate preparation and van der Waals molecular beam epitaxy methods described before.\cite{Chang274,advmat} STM measurements were carried out at 4.7 K on an Unisoku USM1600 system. The Pt-Ir alloy tip was calibrated on the epitaxial silver islands grown on a Si(111) substrate, and d$I$/d$V$ experiments were conducted with a signal recovery lock-in amplifier having a $V_s$ modulation frequency of 913 Hz. { Sample growth and STM studies were performed in the same vacuum system without exposure to air.}

\section{Results and discussion}\label{sec:results}
\subsection{Structure { and STM images} of ferroelectric 2 AL SnTe films}
\begin{figure}
\includegraphics[width=0.46\textwidth]{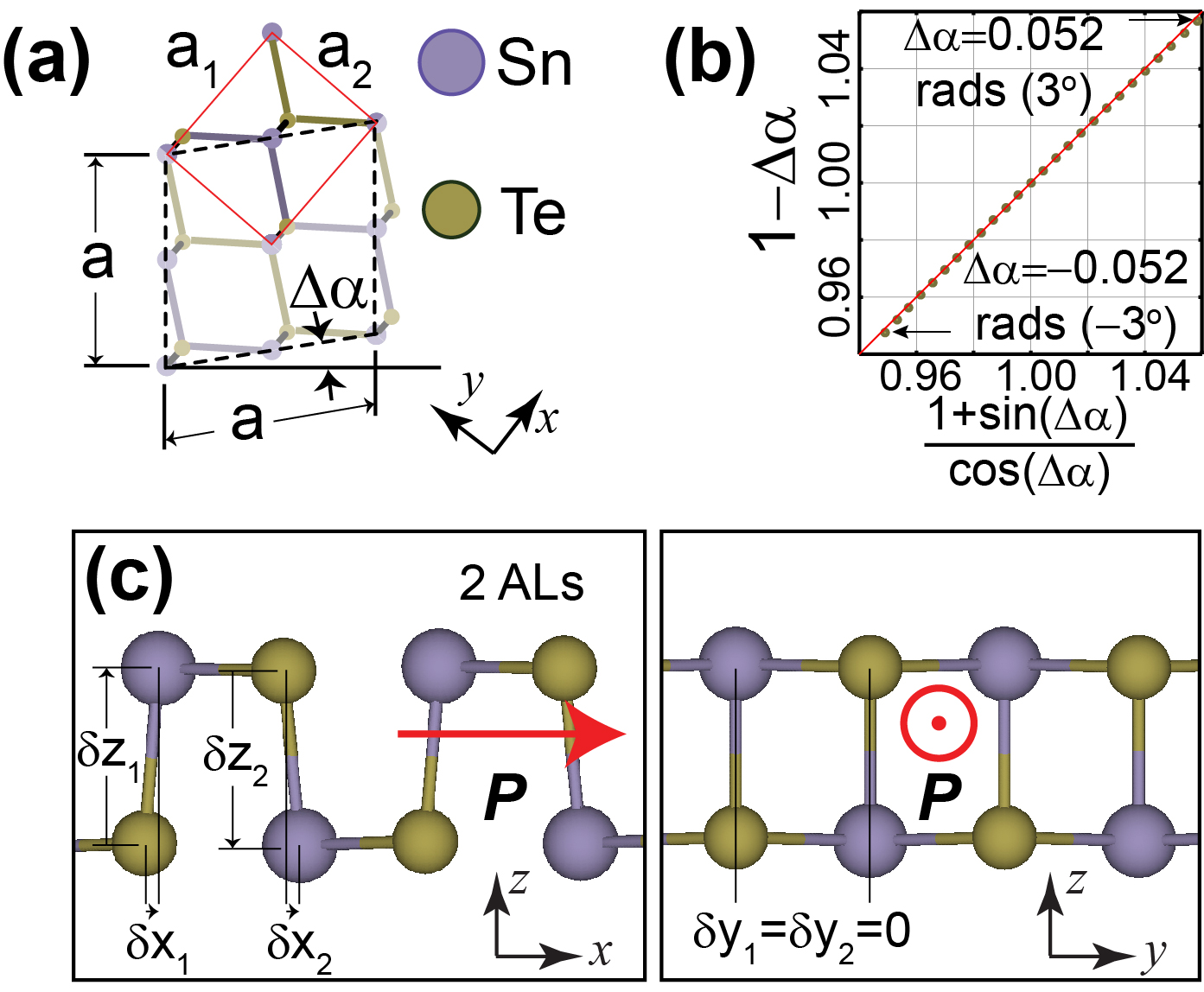}
\caption{(a) Definition of the rhombic distortion angle $\Delta\alpha$. (b) Demonstration of the linear relation among $1+\Delta\alpha$ and $(1+\sin(\Delta\alpha))/\cos(\Delta\alpha)$ up to $\pm 3^{\circ}$. (c) Side views for 2 AL SnTe. Arrows in (c) indicate the spontaneous polarization $\mathbf{P}$ and $\delta x_i$, $\delta y_i$, and $\delta z_i$ ($i=1,2$) are atomic displacements leading to such spontaneous polarization.}\label{fig:fig1}
\end{figure}

The rhombic distortion angle $\Delta \alpha$ in Fig.~\ref{fig:fig1}(a) is related to the orthorhombic in-plane lattice parameters $a_1$ and $a_2$ by:\cite{PhysRevB.97.024110}
\begin{equation}\label{eq:eq1}
\frac{a_1}{a_2}=\frac{1+\sin(\Delta\alpha)}{\cos(\Delta\alpha)}.
\end{equation}
At low temperature and in ultra-thin films, SnTe displays values of $\Delta\alpha$ no larger than $3^{\circ}$ which, as seen in Fig.~\ref{fig:fig1}(b), permits approximating $\Delta\alpha$ as $a_1/a_2-1$ (in radians). Experimentally, the 2 AL SnTe film schematically shown in Fig.~\ref{fig:fig1}(c) registers a value $\Delta \alpha=1.4\pm 0.1^{\circ}$ at 4 K. In our calculations, $a_1=4.728$ \AA\ and $a_2=4.567$ \AA{} for $\Delta \alpha=2.02^{\circ}$. Figure~\ref{fig:fig1}(c) displays a side view of two unit cells of the 2 AL SnTe film, in which the direction of the intrinsic electric dipole $\mathbf{P}$ is explicitly shown.

\begin{figure}[tb]
\includegraphics[width=0.46\textwidth]{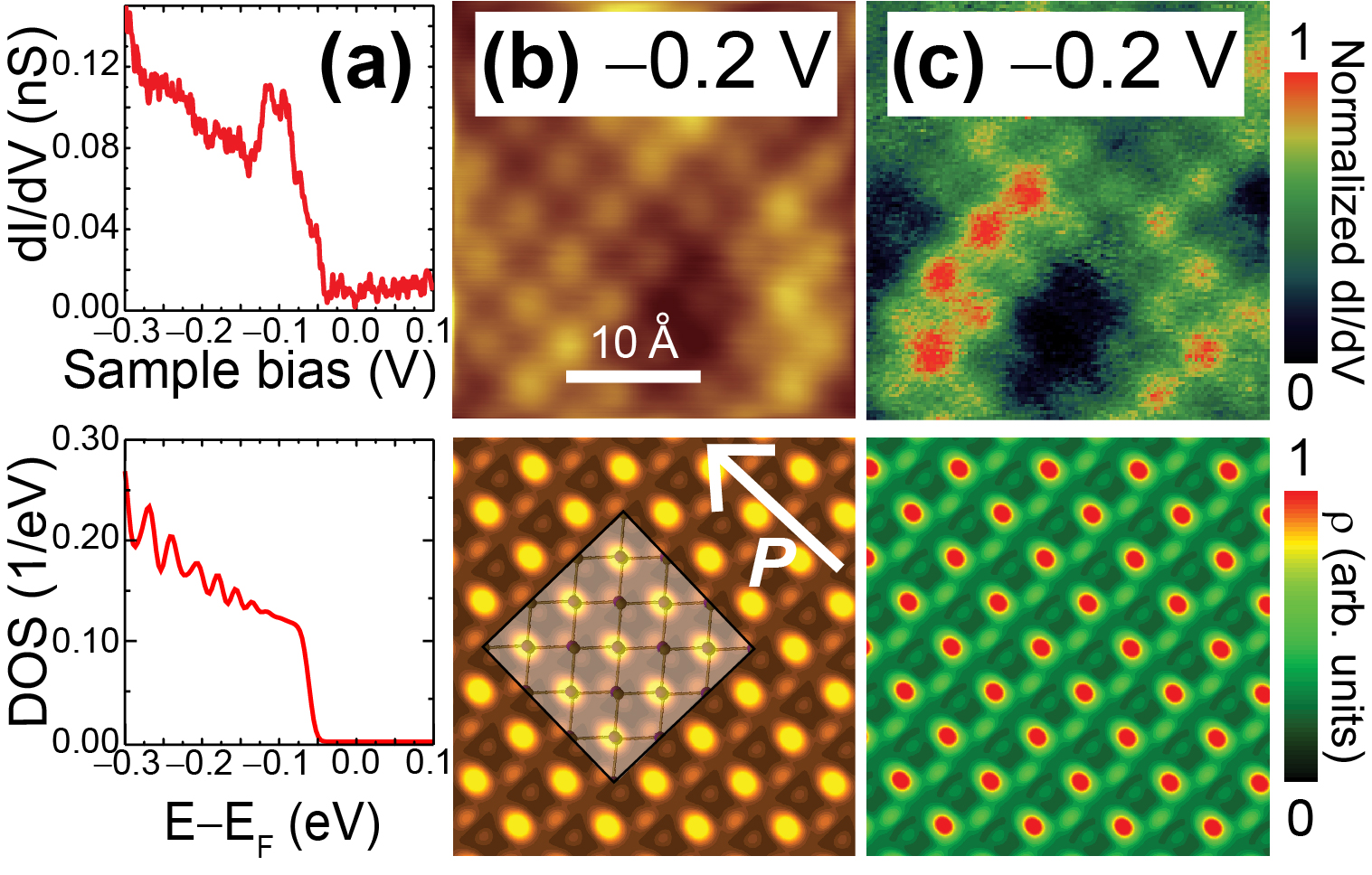}
\caption{(a) Experimental $dI/dV$ curve (upper row) and computed DOS (lower row) { for a 2 AL SnTe film.} The band edge of the DOS curve is aligned with that of the experimental $dI/dV$ curve in order to integrate { the electronic density up to energies consistent with experiment}. (b) STM in topographic mode at a bias of $-0.2$ V (upper row) and setpoint current $I_t = 100$ pA, and its simulated counterpart (lower row; atomic positions of a $2\times 2$ supercell are overlaid there). (c) Upper plot: scanning tunneling spectroscopy (STS) images at $V_s=-0.2$ V, with a setpoint current $I_t = 100$ pA, and a sample bias modulation $V_{mod}= 0.001$ V. Lower subplot shows the electronic density in between $-0.20\pm 0.01$ eV.}\label{fig:fig2}
\end{figure}

\begin{figure*}[tb]
\centering
  \includegraphics[width=0.98\textwidth]{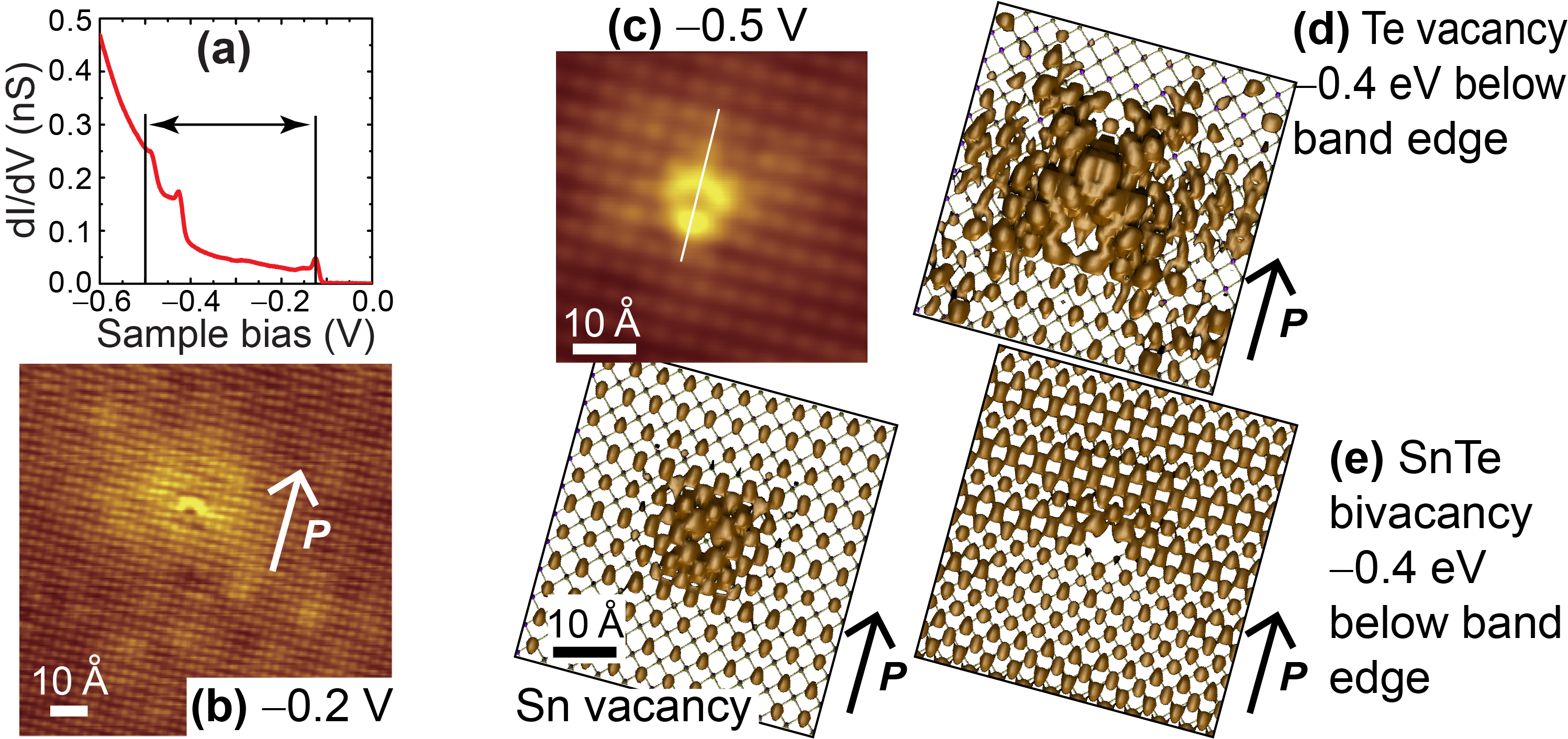}
\caption{(a) $dI/dV$ spectrum away from the bright defect seen in subplot (b). (b) Topographic STM image of a defect at $V_s = -0.2$ V. (c) Upper (lower) plot: experimental (simulated, 3D) topographic image of defect at $-0.5$ V. { (d) Simulated 3D density image of a Te vacancy. (e) Simulated 3D density image of a SnTe bivacancy. The density in subplots (d) and (e) was integrated down to the leftmost vertical solid line shown in subplot (a) for consistency.}}\label{fig:fig3}
\end{figure*}

Experimental $dI/dV$ curves (upper panels) and simulated DOS data (lower panels) are presented for a 2 AL SnTe film { in Fig.~\ref{fig:fig2}. The valence band edge on the DOS has been horizontally displaced to match the experimental band edge, so that simulated STM images are integrated to energies consistent with experiment.}  { The experimental} $dI/dV$ curve in Fig.~\ref{fig:fig2}(a) shows a peak at the band edge which does not appear on the simulated $DOS$, and is likely due to { adatoms} at the graphene/2 AL SnTe interface, { consistent with dark regions on the experimental figure that register a reduced tunneling current.} Aside from that abrupt peak at about $-0.1$ V, the experimental $dI/dV$ and simulated $DOS$ curves can be empirically related by $dI/dV\simeq 0.5\times DOS$.

{ The top subplot in Fig.~\ref{fig:fig2}(b) is an experimental topographic image in which the energy has been integrated down to $-0.2$ V while the lower subplot corresponds to a simulation of the total density from the Fermi energy down to the corresponding experimental energy. Fig.~\ref{fig:fig2}(c), on the other hand, represents the electronic density within a narrow energy range; \emph{i.e.}, the density created by only a handful of electronic wavefunctions.}
Brightest spots that provide the atomic registry in Figs. \ref{fig:fig2}(b,c) are reproduced in simulations. Bright spots in Figs.~\ref{fig:fig2}(b) and \ref{fig:fig2}(c) correspond to the {  exposed Sn sublattice --which protrudes higher than the Te atoms according to Fig.~\ref{fig:fig1}(c)-- as emphasized by an overlaid $3\times 3$ atomistic supercell in Fig.~\ref{fig:fig2}(b). In these plots, the brightest feature in all simulated images looks elongated along the direction parallel to $\mathbf{P}$.}

\subsection{{ On the possible type} of atomic vacancies}
To isolate vacancies and avoid spurious interactions arising from periodic images, a single structural defect (a Sn, Te, or Sn-Te dimer vacancy) was simulated on a 11$\times$11 supercell, { making it computationally expensive to observe these defects on} films thicker than 2 ALs. { Nevertheless, STM images of structural vacancies display high contrast over many atomic sites and have geometrical shapes that ought to be independent of material thickness, giving us confidence that simulations of vacancies on a 2 AL SnTe film do provide relevant information that is representative of vacancies on thicker films.}

Figure~\ref{fig:fig3}(a) displays the experimental $dI/dV$ profile of a SnTe film of unknown thickness. The direction of the intrinsic electric dipole $\mathbf{P}$ was obtained from the band bending at crystal edges (not shown in the Figure) using techniques developed before.\cite{Chang274,advmat} The bright yellow feature in the STM topography image at $V_s = -0.2$ V and $I_t$ = 100~pA in Fig.~\ref{fig:fig3}(b) { will be shown to surround a Sn} vacancy.

In Fig.~\ref{fig:fig3}(c), upper subplot, the topographic feature displayed in Fig.~\ref{fig:fig3}(a) is shown under a smaller bias of $-0.5$ V and at a higher spatial resolution. The bright feature surrounding the dark spot is not radial-symmetric, and its axis of symmetry that is parallel to the arrow indicating the direction of $\mathbf{P}$ in Fig.~\ref{fig:fig3}(b). Figure \ref{fig:fig3}(c), lower subplot, is a simulated 3D isodensity image for a 2 AL SnTe film { on the $3\times 3\times 1$ $k-$point mesh indicated before}; this image was rotated to match the orientation and size of the experimental figure. A spot with no density centered along the Sn vacancy can be seen surrounded by an asymmetric density in the simulated image, { with a larger (smaller) density above (below) the zero-density spot. In addition, the overall size of the simulated feature matches the size of the experimentally seen vacancy related state.} The direction of the electric dipole $\mathbf{P}$ can be directly observed from the simulated structure, and it matches the orientation of polarization determined experimentally.\cite{advmat}

\begin{figure}
\includegraphics[width=0.46\textwidth]{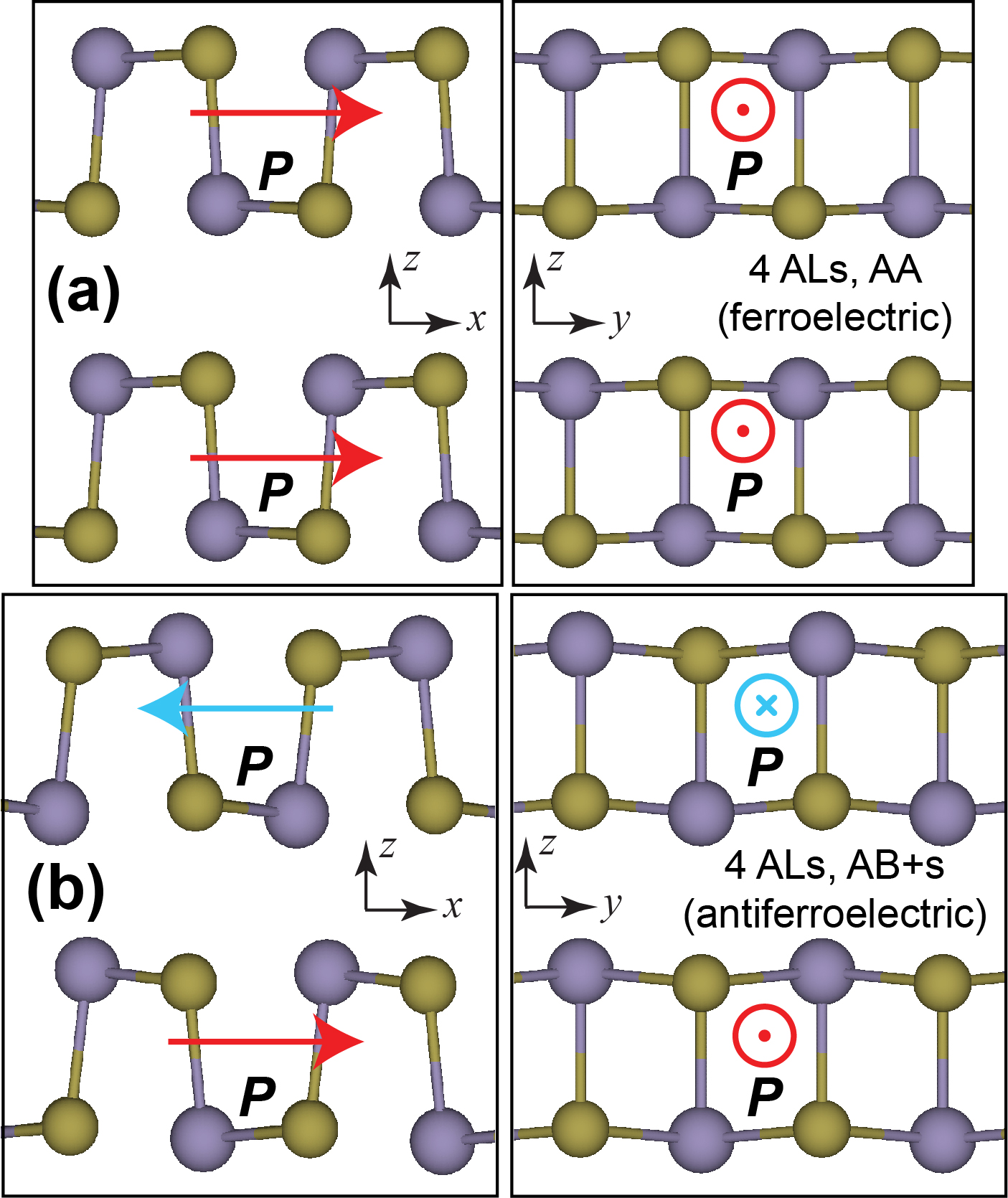}
\caption{(a) 4 ALs with a ferroelectric ($AA$) stacking. (b) 4 ALs with an antiferroelectric stacking, in which the upper 2 AL is shifted by $\mathbf{a}_1/2$ ($AB+s$).  Arrows indicate the spontaneous polarization $\mathbf{P}$ on a given 2 AL.}\label{fig:figNEW}
\end{figure}

\begin{figure}[tb]
  \includegraphics[width=0.49\textwidth]{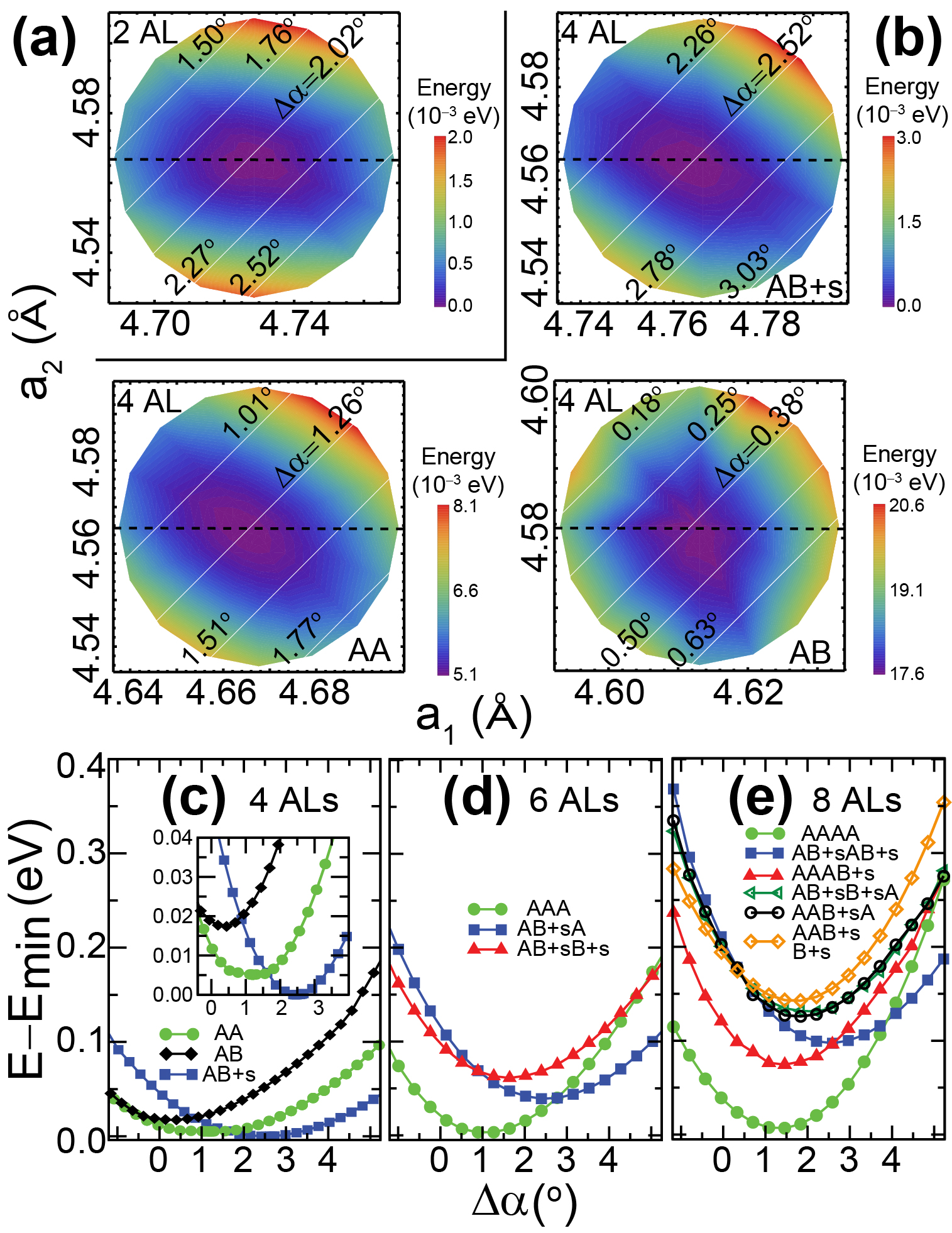}
\caption{Relative energy as a function of $a_1$, $a_2$ for (a) 2 AL SnTe, and (b) 4 AL SnTe in the $AB+s$, $AA$, and $AB$ configurations. Diagonal lines in solid white in subplots (a) and (b) denote a constant value of $\Delta\alpha$. The trends shown in subplot (c) were drawn over constant $a_2$ lines that cross the minimum energy point, shown as dashed lines in subplot (b). (d) and (e): Energy for 6 ALs and 8 ALs drawn over constant $a_2$ lines that include their absolute minimum energies. Every 2 ALs on the 4 AL film are antiferroelectrically coupled, while successive 2AL layers in thicker films are ferroelectrically coupled.}\label{fig:fig4v1}
\end{figure}

{
A Te vacancy, seen in Fig.~\ref{fig:fig3}(d), has a features inconsistent with the experimental STM displayed in Fig.~\ref{fig:fig3}(c) at an energy and isodensity identical to that used in Fig.~\ref{fig:fig3}(c), implying that the experimentally observed feature is not a Te vacancy. Fig.~\ref{fig:fig3}(e) corresponds to the removal of the Sn atom from the upper sub-layer and the Te atom from the lower sub-layer. There, the bright-dark contrast is not as extended as in the case of the single Sn vacancy. In summary, the comparison among experiment and simulations allows us to affirm that the vacancies are due to Sn atoms which also dope the SnTe films with holes.}

\subsection{Antiferroelectrically-coupled 4 AL SnTe films}

The ferroelectric coupling exemplified in Figs. \ref{fig:figNEW}(a) and \ref{fig:figNEW}(b) bears importance on electrostatic energy storage applications because antiferroelectrically-coupled ferroelectrics --such as the structure in subplot \ref{fig:figNEW}(b)-- have been argued to lead to ultra-high-density capacitors.

Up to now, antiferroelectric coupling is induced by substitutional doping,\cite{Xu2017} making it relevant to know whether 4 AL SnTe realizes ferroelectric coupling --where consecutive pairs of 2 ALs have a parallel orientation of their in-plane polarization $\mathbf{P}$, c.f., Fig.~\ref{fig:figNEW}(a)-- or antiferroelectric behavior in which consecutive 2 ALs have antiparallel in-plane polarizations, c.f., Fig.~\ref{fig:figNEW}(b).

There are three complementary experimental tests to determine the ferroelectric coupling of the 4 AL SnTe film:\cite{Chang274} (i) the height profile, (ii) the band bending at the exposed ends, and the magnitude of $\Delta\alpha$ from the Fourier transform of the STM image.
Band bending is larger on a ferroelectrically-coupled (\textit{AA}) 4 AL SnTe film when compared to an antiferroelectrically-coupled (\textit{AB}) 4 AL SnTe film, because the electric field lines cancel out at the exposed edge on the latter case.\cite{Chang274,advmat}

Here, we use energetics and the experimental values of $\Delta\alpha$ for 2 AL and 4 ALs, to demonstrate an antiferroelectric coupling on 4 AL SnTe that is at odds with previous claims of ferroelectric coupling\cite{arxiv1,liu_prl_2018_snte} and consistent with experiment.\cite{Chang274,advmat}

In the present calculations, the $AA$ structure shown in Fig.~\ref{fig:figNEW}(a) has two 2 ALs relatively displaced along the $z-$direction. The structure shown in Fig.~\ref{fig:figNEW}(b) and labeled $AB+s$ (short for $AB$+shift) { has the following coordinates:
\begin{eqnarray*}
\mathbf{b}_1=&(a_1/2+\delta_2,	a_2/2,	z_1)	&\text{  Sn}\\
\mathbf{b}_2=&(\delta_1,	0.0,	0.0)	&\text{  Sn}\\
\mathbf{b}_3=&(0.0, 0.0, z_3)	&\text{  Te}\\
\mathbf{b}_4=&(a_1/2, a_2/2, z_1-z_3)	&\text{  Te}\\
\mathbf{b}_5=&(a_1-\delta_1, a_2/2, z_1+\Delta)	&\text{  Sn}\\
\mathbf{b}_6=&(a_1/2-\delta_2, 0.0, \Delta)	&\text{  Sn}\\
\mathbf{b}_7=&(a_1/2,0.0,z_3+\Delta)	&\text{  Te}\\
\mathbf{b}_8=&(a_1, a_2/2, z_1-z_3+\Delta)	&\text{  Te},
\end{eqnarray*}
with $\delta_1=0.316$, $\delta_2=0.307$, $z_1=3.182$, $z_3=2.967$ and $\Delta=6.179$ (all in \AA) and $a_1$, $a_2$ provided in Table \ref{ta:ta1}.}

The structures shown in Figs.~\ref{fig:fig1} and \ref{fig:figNEW} are the result of a full structural optimization using a dense meshing procedure in which $a_1$ and $a_2$ are preset to fine mesh locations, and only atomic positions are relaxed. This procedure yields the energy {\em versus} $a_1$, $a_2$ plots shown in Fig.~\ref{fig:fig4v1}(a), in which iso-$\Delta\alpha$ lines { oriented at 45$^{\circ}$ resulting from Eqn.~\ref{eq:eq1}} are displayed as well. The optimal lattice constants and relative energies are reported in Table \ref{ta:ta1} for each of these structures too. In this Table, the labels $AA$, $AAA$ and $AAAA$ stand for ferroelectric coupling.

\begin{table}[tb]
 \caption{Lattice parameters, $\Delta\alpha$, and relative energies of ultra-thin freestanding SnTe films. Experimentally, $\Delta\alpha(4AL)>\Delta\alpha(2 AL)$.\cite{Chang274} The values reported for 2, 4 $AB+s$, 6 and 8 ALs correspond to ground-state structures.}\label{ta:ta1}
\centering
\begin{tabular}{|c|cc|cc|c|}
\hline
\hline
Structure & $a_1$ (\AA) & $a_2$ (\AA) & $\Delta\alpha$           &$\Delta\alpha$& Energy\\
          &             &             &  theo. (${}^{\circ}$)  &expt. (${}^{\circ}$)& diff. (eV)\\
\hline
\hline
2 AL     & 4.728 & 4.567 & $2.02$  &$1.4\pm0.1$  &\\
\hline
\hline
4 AL, $AA$      & 4.668 & 4.566 & 1.28 &                                    & 0.0145\\
\hline
4 AL, $AB$      & 4.662 & 4.565 & 1.22 &                                    & 0.0150\\
\hline
4 AL, $AB+s$        & 4.766 & 4.565 & 2.52 &$1.9\pm0.1$ & 0.0000\\
\hline
\hline
6 AL, $AAA$     & 4.656 & 4.563& 1.18 & &\\ 
\hline
\hline
8 AL, $AAAA$    & 4.651 & 4.564& 1.09 & &\\ 
\hline
\hline
\end{tabular}
\end{table}

\begin{figure}[tbp]
  \includegraphics[width=0.480\textwidth]{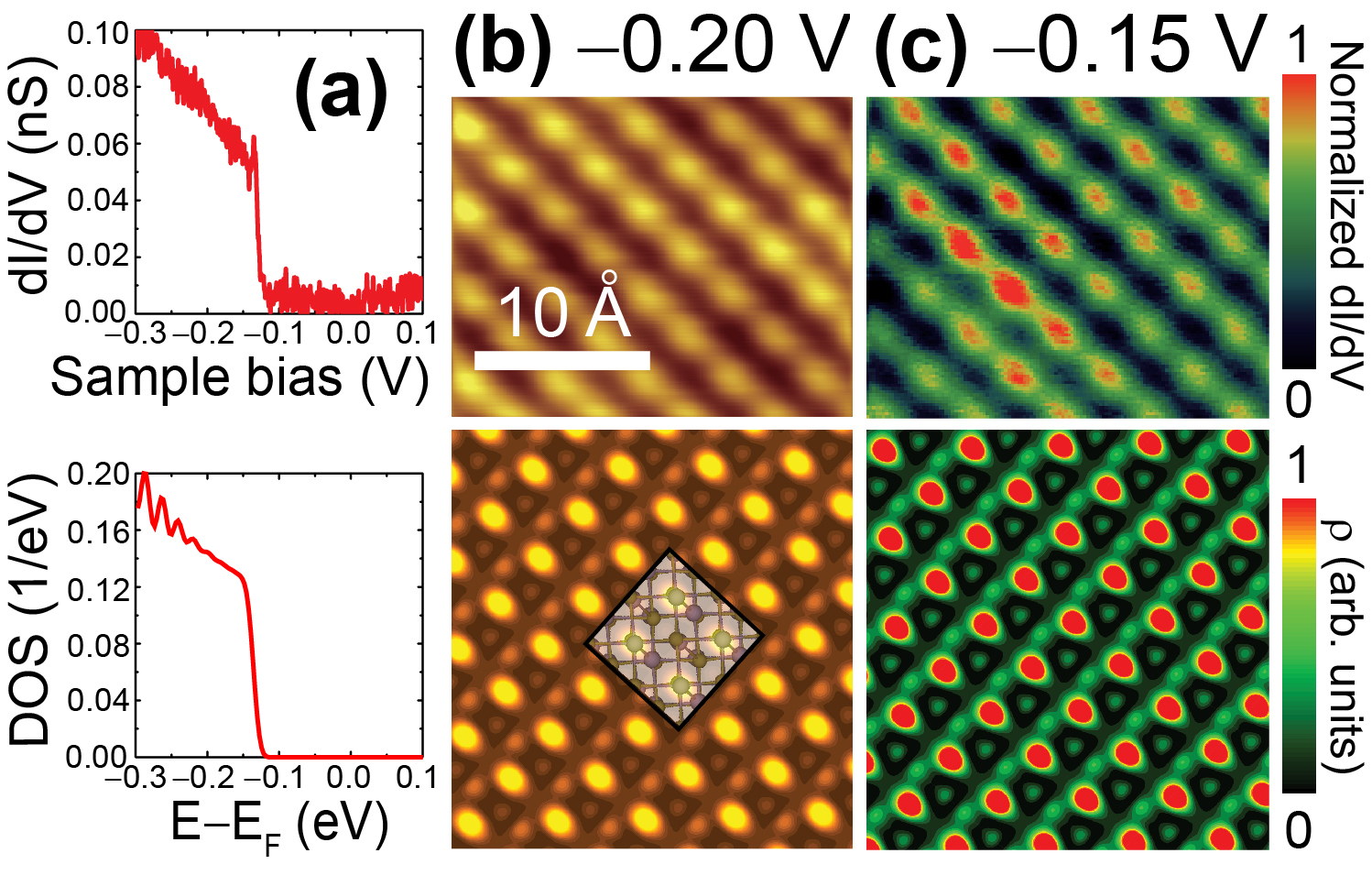}
\caption{(a) d$I$/d$V$ and $DOS$ for a 4 AL SnTe film. The simulated structure has an $AB+s$ (antiferroelectric) structure. (b) Experimental (upper row) and simulated (lower row) STM images at a bias voltage of $-0.2$ V. (c) Upper plot: scanning tunneling spectroscopy (STS) images at $V_s=-0.15$ V, with a setpoint current $I_t = 100$ pA, and a sample bias modulation $V_{mod}= 0.001$ V. Lower subplot shows the electronic density in between $-0.15\pm 0.01$ eV.}\label{fig:fig4}
\end{figure}

{ Figure \ref{fig:fig4v1}(c) provides one-dimensional plots that cut across the $\Delta\alpha$ paths that cross the absolute minima in Fig.~\ref{fig:fig4v1}(b). These plots provide a comparative study of energetics \emph{versus} the relative placement of consecutive 2 ALs.} The important point from Fig.~\ref{fig:fig4v1} is that the lowest-energy 4 AL structure is antiferroelectrically coupled.

\begin{figure*}[tb]
\includegraphics[width=0.98\textwidth]{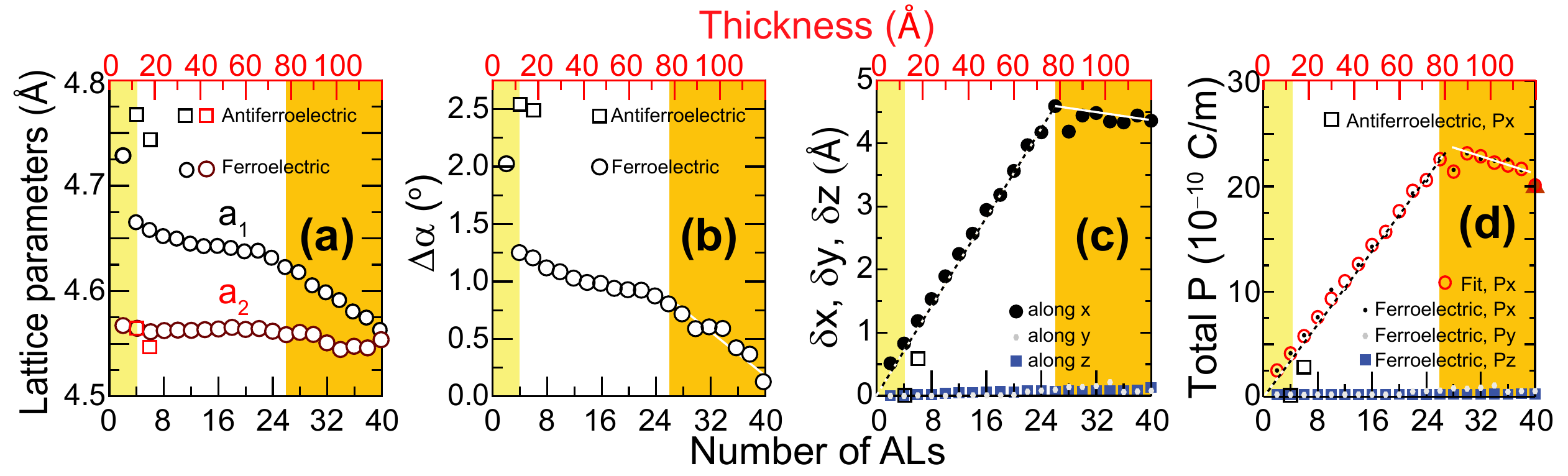}
\caption{Thickness dependence of (a) the in-plane lattice parameters $a_1$ and $a_2$, (b) the rhombic distortion angle $\Delta\alpha$, (c) the structural distortions $(\delta_x,\delta_y,\delta_z)$, and (d) the total electric dipole $\mathbf{P}$ for SnTe films from 2 to 40 ALs. Circles correspond to a ferroelectric ($AA$) coupling among consecutive 2 ALs, while open squares describe slabs with an antiferroelectric, $AB+s$ coupling. Yellow, white and orange colors highlight three trend regions.}
\label{fig:fig6}
\end{figure*}

Fig.~\ref{fig:fig4} displays experimental (upper row) and computational (lower row) results supporting the antiferroelectric coupling of 4 AL SnTe.\cite{advmat} There, the empirical relation $dI/dV\simeq 0.5\times DOS$ (established in Fig.~\ref{fig:fig2}(a)) can be seen again. Furthermore, the location of bright spots between experimental and simulated STM images agree in subplots \ref{fig:fig4}(b) and \ref{fig:fig4}(c). Despite of the increased spatial resolution of the simulated STM image at $-0.15$ V when compared with its experimental counterpart, Fig.~\ref{fig:fig4}(c), \emph{the dark diagonal feature can be observed along the elongated direction ($a_1$) in both images}. Following Equation \ref{eq:eq1}, such relative elongation of $a_1$ with respect to $a_2$ that leads to the asymmetric dark diagonal lines at $-45^{\circ}$ is necessary to achieve the value experimental value of $\Delta\alpha$.

In conclusion, experiment\cite{Chang274,advmat} and the present calculations confirm an antiferroelectric coupling of 4 AL SnTe films, while Table \ref{ta:ta1} and Fig.~\ref{fig:fig4v1} indicate that these unsupported films with more than 4 ALs are ferroelectrically coupled.

\subsection{Atomistic structure and electronic bandstructures of thicker films}

The antiferroelectric coupling of 4 AL SnTe and the ferroelectric coupling on the rhombic bulk discussed thus far imply the existence of a critical thickness at which the antiferroelectrically-coupled thin films transition onto a (bulk-like) ferroelectrically coupled SnTe.

Previous observation invites to examine the ferroelectric coupling of thicker films following the bottom-up approach pursued thus far. To this end, and as reported in Figs. \ref{fig:fig4v1}(d) and \ref{fig:fig4v1}(e), 6 AL SnTe slabs and 8 AL SnTe freestanding slabs were first considered, to find that ferroelectric coupling was preferred in both instances. For that reason, all thicker films studied here were stacked in a ferroelectric fashion consistent with bulk SnTe.
Reference ~\onlinecite{Stroscio} indicates that films with a thickness in excess of 100 \AA{} --corresponding to about 36 ALs-- behave as bulk SnTe. Using such experimental guidance to set an upper thickness limit, the thickness dependence of in-plane lattice parameters $a_1$ and $a_2$ and $\Delta\alpha$ for SnTe films ranging from 2 to 40 ALs {  as obtained computationally} are shown in Figs.~\ref{fig:fig6}(a) and \ref{fig:fig6}(b), respectively. Data in circles in Fig.~\ref{fig:fig6} corresponds to a ferroelectric coupling among consecutive 2 ALs, while squares at the left ends of these plots describe slabs with an antiferroelectric $AB+s$ coupling. The similar trends in between $a_1$ and $\Delta \alpha$ in Figs.~\ref{fig:fig6}(a) and \ref{fig:fig6}(b) arise from the linear dependency of $\Delta\alpha$ on $a_1/a_2$ in Eqn.~\ref{eq:eq1} and the almost-constancy of $a_2$ in Fig.~\ref{fig:fig6}(a).

There is an abrupt increase on $a_1$ and $\Delta\alpha$ in going from 2 AL to 4 ALs, as the $AB+s$ phase increases $a_1$ (squares in Figs.~\ref{fig:fig6}(a)), at the expense of reducing its dipole moment down to zero in \ref{fig:fig6}(d). The area in white in Fig.~\ref{fig:fig6}(a) shows an almost constant $a_2$, and a decay of $a_1$ by 0.0013 \AA{} per AL, making $\Delta\alpha$ in Fig.~\ref{fig:fig6}(b) decay by  0.025$^{\circ}$ per AL as the SnTe film gradually turns into a bulk structure. In between 30 and 40 ALs, the decay of $a_1$ in Fig.~\ref{fig:fig6}(a) becomes more drastic (0.0042 \AA{} and 0.050$^{\circ}$ per AL, respectively), so that $\Delta\alpha\sim 0.1^{\circ}$ at 40 ALs in Fig.~\ref{fig:fig6}(b).

Atomistic displacements $\delta_x$, $\delta_y$ and $\delta_z$ among anion and cations on a given 2 AL shown in Fig.~\ref{fig:fig1} can be linked to the total polarization observed in these films. In order to characterize atomistic displacements for thicker films we define:
\begin{equation}\label{eq:2}
\delta x\equiv\sum_{n=1}^{N/2}\sum_{i=1}^2\delta x_{i,n},
\end{equation}
as the sum of displacements along the $x-$directions for thicker slabs, where $N$ is the number of ALs on a given slab (there are two displacements per 2 AL as seen in Fig.~\ref{fig:fig1}), with similar expressions for $\delta y$ and $\delta z$. $\delta x$, $\delta y$ and $\delta z$ are displayed in Fig.~\ref{fig:fig6}(c). There, $\delta x$ increases by 0.17 \AA{} per AL up to 26 ALs, and then slightly decreases for thicknesses in between 28 and 40 ALs as a result of the sudden compression of $a_1$ seen in Fig.~\ref{fig:fig6}(a). In turn, $\delta y$ and $\delta z$ remain equal to zero.

\begin{figure*}[tb]
\includegraphics[width=0.98\textwidth]{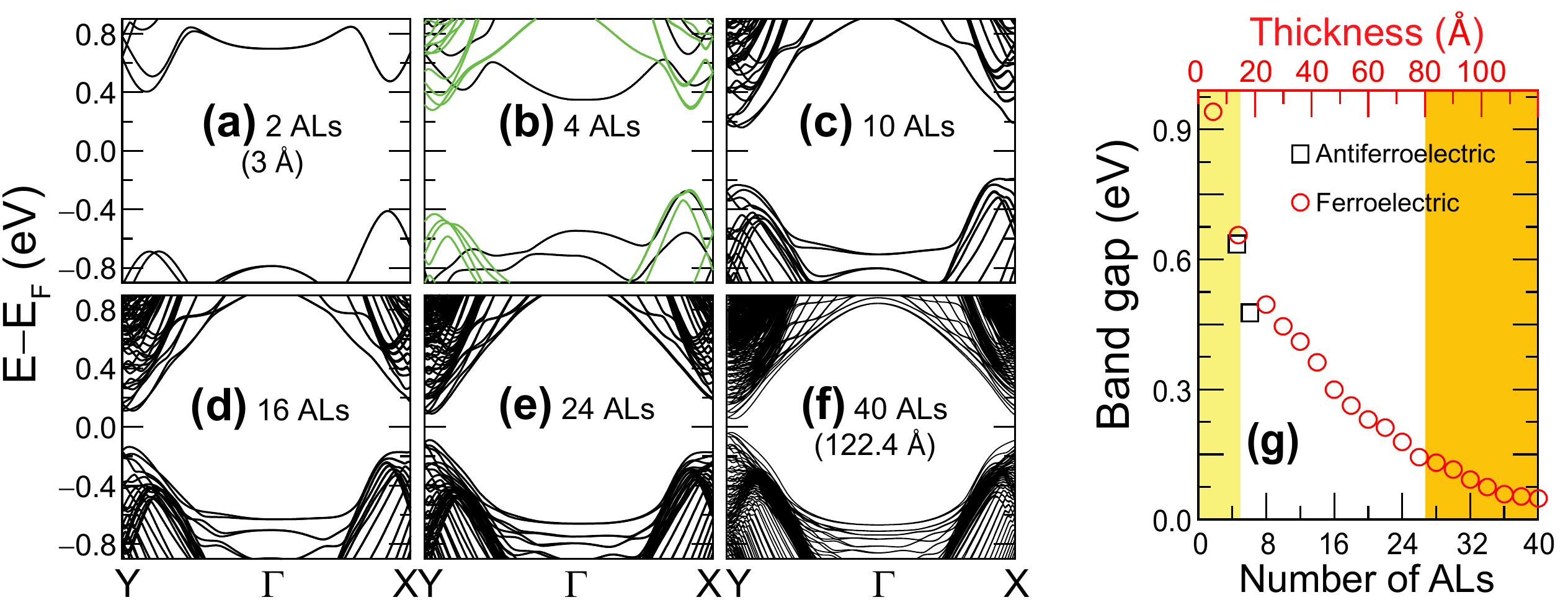}
\caption{Thickness dependence of the electronic band structure of SnTe ultra-thin films of selected thicknesses: (a) 2 ALs. (b) 4 ALs (green bands correspond to the AB+s phase, and black bands  to the AA phase. (c) 10 ALs, (d) 16 ALs, (e) 24 ALs, and (f) 40 ALs. Plots (a-f) were obtained with spin-orbit coupling (SOC). (g) Magnitude of the electronic band gap {\em versus} the number of ALs.}
\label{fig:fig8}
\end{figure*}

In Fig.~\ref{fig:fig6}(d), we obtained the electric dipole $\mathbf{P}$ of a 2 AL slab using the standard Berry phase approach,\cite{Berry} and linked $\mathbf{P}$ to the magnitude of $\delta x$. (We did so using {\em VASP}, with structures obtained from the {\em SIESTA} code.) This permitted adding the module on the standard Berry phase estimation to a periodic term that was consistent with the magnitude of $\delta x$. The total polarization increases by $0.9\times 10^{-10}$ $C/m$ per AL up to 26 ALs. From then on, both $\delta_x$ and $P_x$ decrease at a rate of 0.02 \AA{} per AL and $0.4 \times 10^{-10}$ $C/m$ per AL up to 40 ALs. The sudden drop of $a_1$ past 30 ALs does reduce the overall magnitude of $\delta x$ despite of the subsequent addition of MLs, but it never brings the dipole all the way to zero for what it would be a rocksalt conformation. Instead, the reported non zero dipole is linked to the rhombic nature of films containing more than 4 ALs.

Lastly, we display the electronic structure with spin-orbit coupling (SOC) turned on of SnTe slabs with increasing thicknesses in Figs. \ref{fig:fig8}(a-f). Figure~\ref{fig:fig8}(a) displays the 2 AL SnTe thin film as an indirect bandgap semiconductor.\cite{Gomes} As seen in Fig.~\ref{fig:fig8}(b), such indirect bandgap remains for 4 AL SnTe with $AA$ (black lines) and $AB+s$ stacking (green lines)  due to the largely dissimilar magnitudes of $a_1$ and $a_2$ in both structures. The indirect band gap persists up to 8 ALs.

Even though the band structure is not symmetric around the $X-$ and $Y-$points in Fig.~\ref{fig:fig8}(c), the band gap turns direct for a thickness of 10 ALs. Seeing the full sequence of subplots, Figs. \ref{fig:fig8}(a-f), the band gap reduces its value as the thickness increases. This is emphasized by showing its magnitude in Fig.~\ref{fig:fig8}(g).

\section{Conclusion}\label{sec:conclusions}

In conclusion, { and despite of its longevity, SnTe remains an important material in Condensed Matter Physics,} and the structural evolution of SnTe from 2 to 40 ALs has been provided here. 4 AL SnTe favors an antiferroelectric coupling, while suspended films with thicknesses ranging from 6 to 40 AL were ferroelectrically-coupled. The evolution of the rhombic distortion angle, the electric polarization, and of the electronic band structure have been provided as well. The atomistic structures and resulting electric dipole moments and electronic bandstructures are found to be different from those obtained by capping bulk structures, especially for ultrathin films. The information provided here is expected to better understand the coupling among atomistic structure and the fascinating material properties of SnTe.

\begin{acknowledgments}
Research at Arkansas was supported by the U.S. Department of Energy, Office of Basic Energy Sciences, Division of Materials Science and Engineering under Early Career Award  DE-SC0016139. Calculations were performed at the Center for Nanoscale Materials at Argonne National Laboratory. K.C. and S.S.P.P. were funded by the Deutsche Forschungsgemeinschaft (DFG, German Research Foundation) - Project number PA 1812/2-1. K.C., Q.-K.X., X.C. and S.-H.J. were supported by National Natural Science Foundation of China (Grant No. 51561145005) and Ministry of Science and Technology of China (Grant No. 2016YFA0301002).

\end{acknowledgments}

%

\end{document}